\documentclass[]{elsarticle}

\usepackage{lineno,hyperref}
\usepackage{amsmath}
\modulolinenumbers[5]
\usepackage{subfigure}
\usepackage{xcolor}
\usepackage{placeins}
\usepackage{caption}
\usepackage{subcaption}
\usepackage{xr}
\journal{Journal}

\graphicspath{{figs/}}

\begin{document}

\begin{frontmatter}

\title{Learning in ensembles of proteinoid microspheres}


\author{Panagiotis Mougkogiannis*}
\author{Andrew Adamatzky}
\address{Unconventional Computing Laboratory, UWE, Bristol, UK}

\cortext[cor1]{Corresponding author: Panagiotis.Mougkogiannis@uwe.ac.uk (Panagiotis Mougkogiannis)}

\begin{abstract}
Proteinoids are thermal proteins which form microspheres in water in presence of salt. Ensembles of proteinoid microspheres exhibit passive non-linear electrical properties and active neuron-like spiking of electrical potential. We propose that various neuromorphic computing architectures can be prototyped from the proteinoid microspheres. A key feature of a neuromorphic system is a learning. 
Through the use of optical and resistance measurements, we study mechanisms of learning in ensembles of proteinoid microspheres. 
We anlyse 16 types of proteinoids, study their intrinsic morphology and electrical properties. We demonstrate that  proteinoids can learn, memorize, and habituate, making them a promising candidate for novel computing. 
\end{abstract}

\begin{keyword}
   thermal proteins \sep proteinoids \sep microspheres \sep unconventional computing
\end{keyword}

\end{frontmatter}


\section{Introduction}

Proteinoids, or thermal proteins, are produced by heating amino acids to their melting point and initiation of polymerisation to produce polymeric chains, which are swelled in aqueous solution forming hollow structures known as a microspheres~\cite{harada1958thermal,fox1992thermal}. Being able to form spontaneously from simple chemical mixes, proteinoids are speculated to have played a crucial role in the early phases of Earth's creation of life~\cite{harada1958thermal,fox1992thermal,fiore2019origin}. A remarkable feature of proteinoids, which earned them name of proto-neurons~\cite{bi1994evidence, fox1992thermal,fox1995experimental}, is that the microspheres maintain a membrane potential 20~mV to 70~mV without any stimulating current~\cite{przybylski1985excitable} and exhibit oscillations of the electrical potential strikingly similar to action potential of neurons~\cite{ishima1981electrical,przybylski1982membrane}. These oscillations can be observed for several days and even weeks~\cite{przybylski1985excitable,bi1994evidence}. In \cite{adamatzky2021towards} we proposed to use ensembles of the proteinoid microspheres as neuromorphic unconventional computing devices. A key feature of a neuromorphic system is a learning. A learning in physical systems is a formation of conductive pathways~\cite{xu2018effect}. In order to trigger the creation of conducting pathways, the proteinoid network must be subjected to electrical stimuli, e.g. by applying a potential difference between two electrodes inserted in the ensemble of proteinoid microspheres. The voltage should lead reorganization of the proteinoid molecules and generating preferrable routes for the flow of electrical current. Rearrangements of proteinoids molecules might create structures morphologically similar to  neural dendrites and axons paths~\cite{lico2023biological,glock2021translatome,ludwig2022neuroanatomy,paus2022tracking,bromberg2022quantifying}.  

\begin{figure}[!tbp]
\centering
\includegraphics[width=1\textwidth]{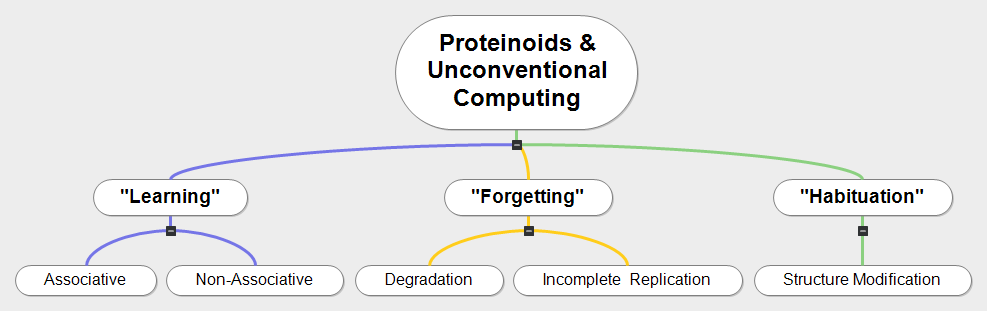}
\caption{ The scheme illustrates how proteinoids can be used in unconventional computing, with a specific emphasis on their capacity for learning and memory, as well as their habituation to stimuli and the ability to adapt their behaviour accordingly. Proteinoid degradation, imperfect replication, structural alteration, and incomplete replication are all proven to play a role in the ``learning'', ``forgetting'', and ``habituation'' processes.
}
\label{pppssssss9vghj99}
\end{figure}

We look into the conductivity and habituation properties of proteinoids to determine their potential in electronics. Fig.~\ref{pppssssss9vghj99} emphasises the three important parameters of learning, forgetting, and habituation, which show the potential of proteinoids in unconventional computing~\cite{tayur2021unconventional,ziegler2020novel}. In particular, Fig.~\ref{pppssssss9vghj99} demonstrates how proteinoids can be employed to keep track of data and adjust to new conditions.

Superior efficiency, reduced power consumption, and increased connectivity with other devices are just some of the benefits that can be realised by using novel semiconductor materials instead of more traditional options. Third-generation semiconductor materials based on SiC and GaN, and two-dimensional semiconductor materials based on graphene, are two examples of innovative semiconductor materials~\cite{stahl2003emergence}.
Proteinoids are materials with the ability to ``learn" from their surroundings, ``forget" and ``habituate," and alter their resistance accordingly. We will investigate how innovative semiconductors can benefit from these features to improve their performance and reliability.\par

Associative learning is a form of learning that takes place when a novel semiconductor learns to link an input with a stimulus~\cite{li2022account,tan2020monadic,li2022account}. By recognising and responding to patterns of input and output, proteinoids can fine-tune their resistance. An innovative semiconductor, for instance, might be programmed to increase its resistance in response to stressors like high temperatures and humidity.

When proteinoids are subjected to the same stimuli over and over again, a sort of learning known as ``non-associative" learning takes place~\cite{mondal2022all,Ioannou2018,Non}. A proteinoid might, for instance, be programmed to lower its resistance in response to particular sounds or frequencies. Proteinoids can employ this form of learning to adjust to new environments.
Proteinoids can experience several forms of forgetting~\cite{berry2012dopamine} depending on how they lose their training data: degradation or incomplete replication. This forgetting can be exploited to reduce the robustness and lifespan of new semiconductors in many settings. An example would be instructing a proteinoid to forget what it has learnt after a particular period of time has passed. Useful for preventing new semiconductors from being weighed down by antiquated data, this form of forgetting can be implemented.

To break down a proteinoid into smaller molecules, a certain amount of voltage, known as the ``critical voltage," must be applied~\cite{Criticalsgas}. A proteinoid's composition and structure establish its critical resistance and critical voltage. Proteinoids rich in sulfur-containing amino acids tend to have higher critical resistance and critical voltage than those rich in polar amino acids. Greater molecular weight proteinoids also have higher critical resistance and voltage values than smaller molecular weight proteinoids.

The development of proteinoid-powered devices, such as biosensors and unconventional computing, increases the significance of this discovery. To elucidate the temporal behaviour of proteinoids, the critical resistance and voltage of the proteinoid network will be examined. This study also offers information on the connection between the structure of the proteinoid network and its electrical conductivity. The findings of this study can be utilised to the creation of proteinoid networks with improved functionality.

\section{Methods}
\label{methods}

L-glutamic acid (L-Glu), L-aspartic acid (L-Asp), L-histidine (L-His), L-lysine acetate (L-Lys), L-phenylalanine (L-Phe), and L-arginine (L-Arg) with a purity of greater than 98\% (Sigma Aldrich) and poly(L-lactic acid) from (Polysciences) have been used. The gold coated proteinoids were examined using an FEI Quanta 650 field emission scanning electron microscopy, and all chemicals employed were used as is. Imaging samples that have been gold-coated prior to imaging will produce higher-quality results by minimising the negative impact of charging effects on the images. Proteinoids have been prepared and characterised in a similar fashion to what has been described previously in~\cite{mougkogiannis2023low}. 

The absorbance of proteinoids at room temperature was determined using a UV-visible spectrometer (Jenway 7200). The recommended calibration range of 200-800~nm was employed, and the device was calibrated accordingly. Absorbance data for the proteinoids was measured at 0.5~nm intervals.

A Keithley 2450 sourcemeter was used to measure the electrical properties of the proteinoids. All readings were taken at room temperature of 19-20$^o$C. 

Matlab was used to analyse and plot the data from the UV visible spectrometer and the source metre. We looked for relationships between the absorbance of proteinoids and their current, voltage, and resistance by comparing these quantities.

\section{Results}

\subsection{Structural Characteristics and Optical Band Gap}

\begin{figure}[!tbp]
\centering
\includegraphics[width=1\textwidth]{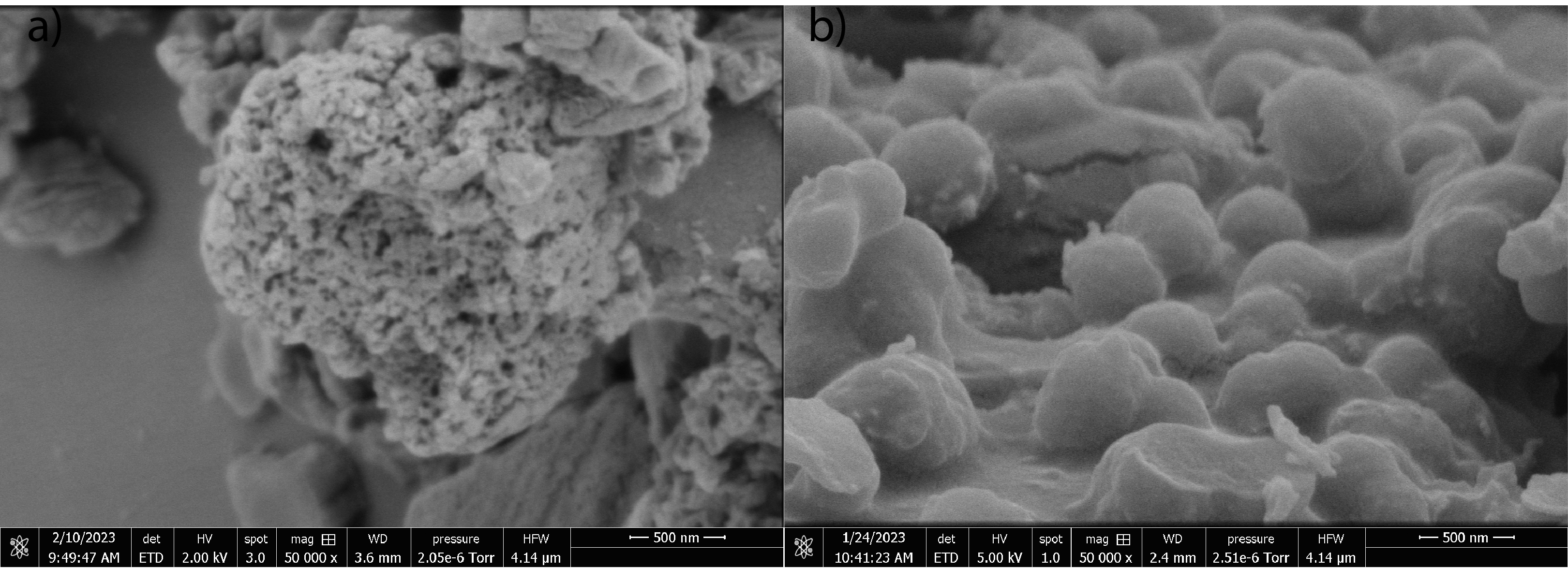}
\caption{ Proteinoids a) L-Phe:L-Arg and b) L-Glu:L-Phe:L-Asp appear as spherical structures in this scanning electron microscope picture.
}
\label{ppp999}
\end{figure}

Proteinoids of L-Phe:L-Arg and L-Glu:L-Phe:L-Asp formed spherical microspheres, as seen in scanning electron microscopy (SEM) pictures (Fig.~\ref{ppp999}). This occurs as a result of the assembly of a polypeptide structure from amino acids, which mimics the function of proteins.

\begin{figure}[!tbp]
\centering
\subfigure[]{\includegraphics[width=0.4\textwidth]{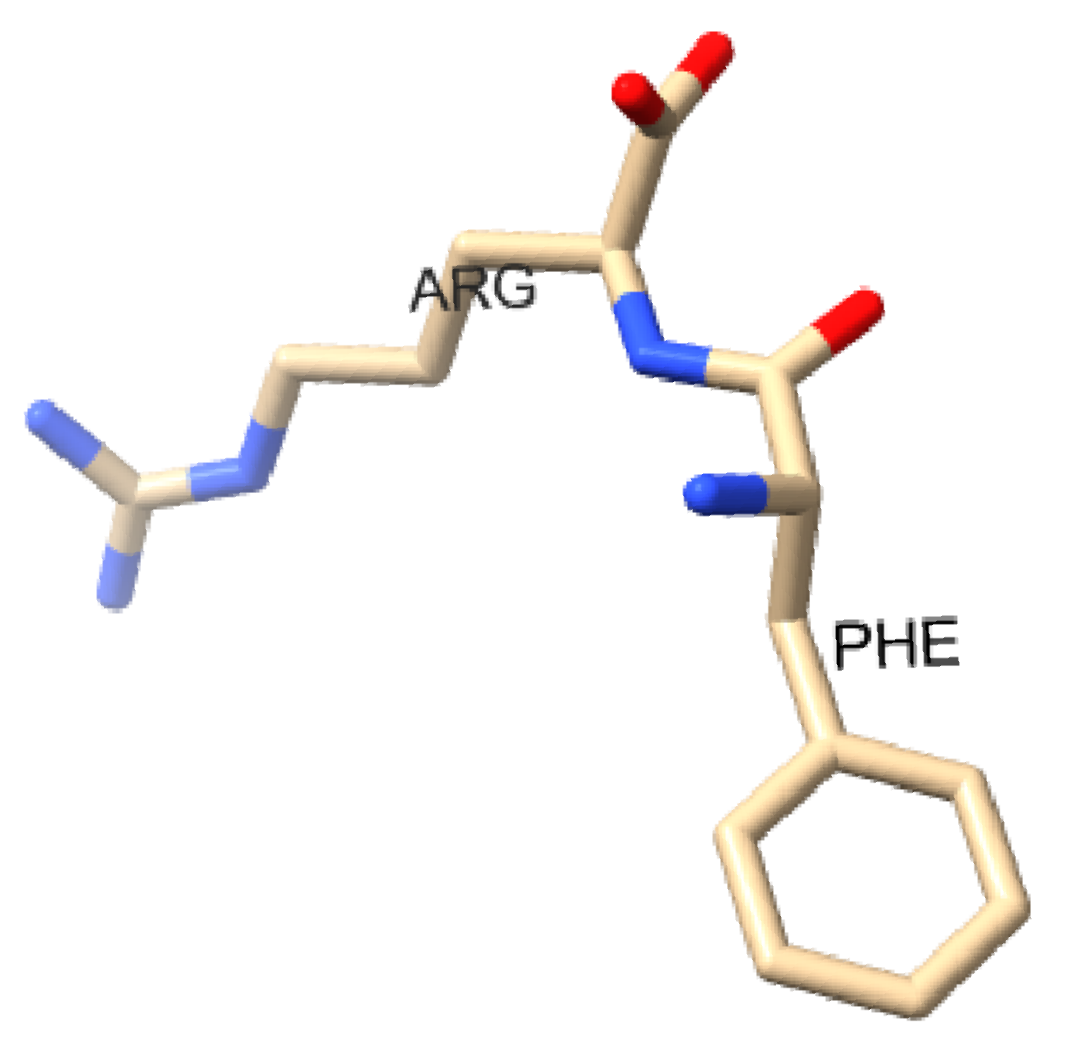}}
\subfigure[]{\includegraphics[width=0.59\textwidth]{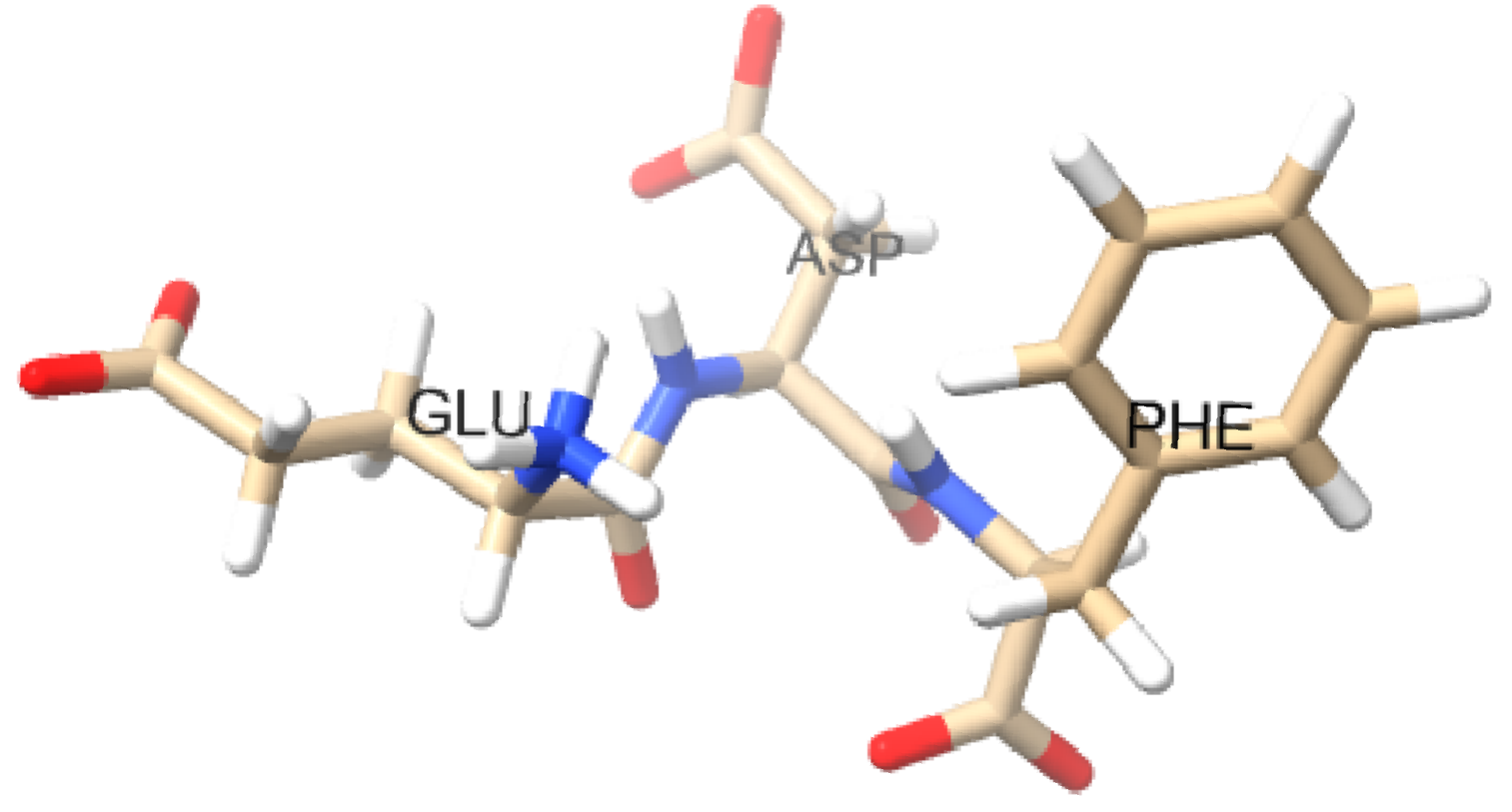}}
\caption{ a) L-Phe:L-Arg and b) L-Glu:L-Phe:L-Asp proteinoids, with nitrogen shown in blue, oxygen in red, and carbon in yellow, respectively. The L-Phe:L-Arg structure is made up of just those two amino acids, while the L-Glu:L-Phe:L-Asp structure adds glutamic acid to the scene. Structures generated in Chimera.
}
\label{ppp9vghj99}
\end{figure}

The polypeptide structure is a stable helical form held together by the strand due to the hydrogen bonds between amino acids. Due to the strong interactions between amino acid side chains, proteinoids take the shape of a sphere when they adopt this helical conformation. Two proteinoids, L-Phe:L-Arg and  L-Glu:L-Phe:L-Asp, have their molecular structures depicted in Fig.~\ref{ppp9vghj99}. 

The precise arrangement of amino acids in the proteinoid is also a contributing factor in the production of perfectly round microspheres. Combinations of aromatic (phenylalanine) and charged (arginine and aspartic acid) amino acids can be found in proteinoids like L-Phe:L-Arg and L-Glu:L-Phe:L-Asp. In contrast to the hydrophilic regions formed by arginine and aspartic acid, which are attracted to water molecules, the hydrophobic regions formed by the aromatic amino acid phenylalanine surround the proteinoids. By balancing out the hydrophilic and hydrophobic regions, the proteinoids take a spherical shape due to the overall hydrophobic surface tension.

\begin{figure}[!tbp]
\centering
\includegraphics[width=1\textwidth]{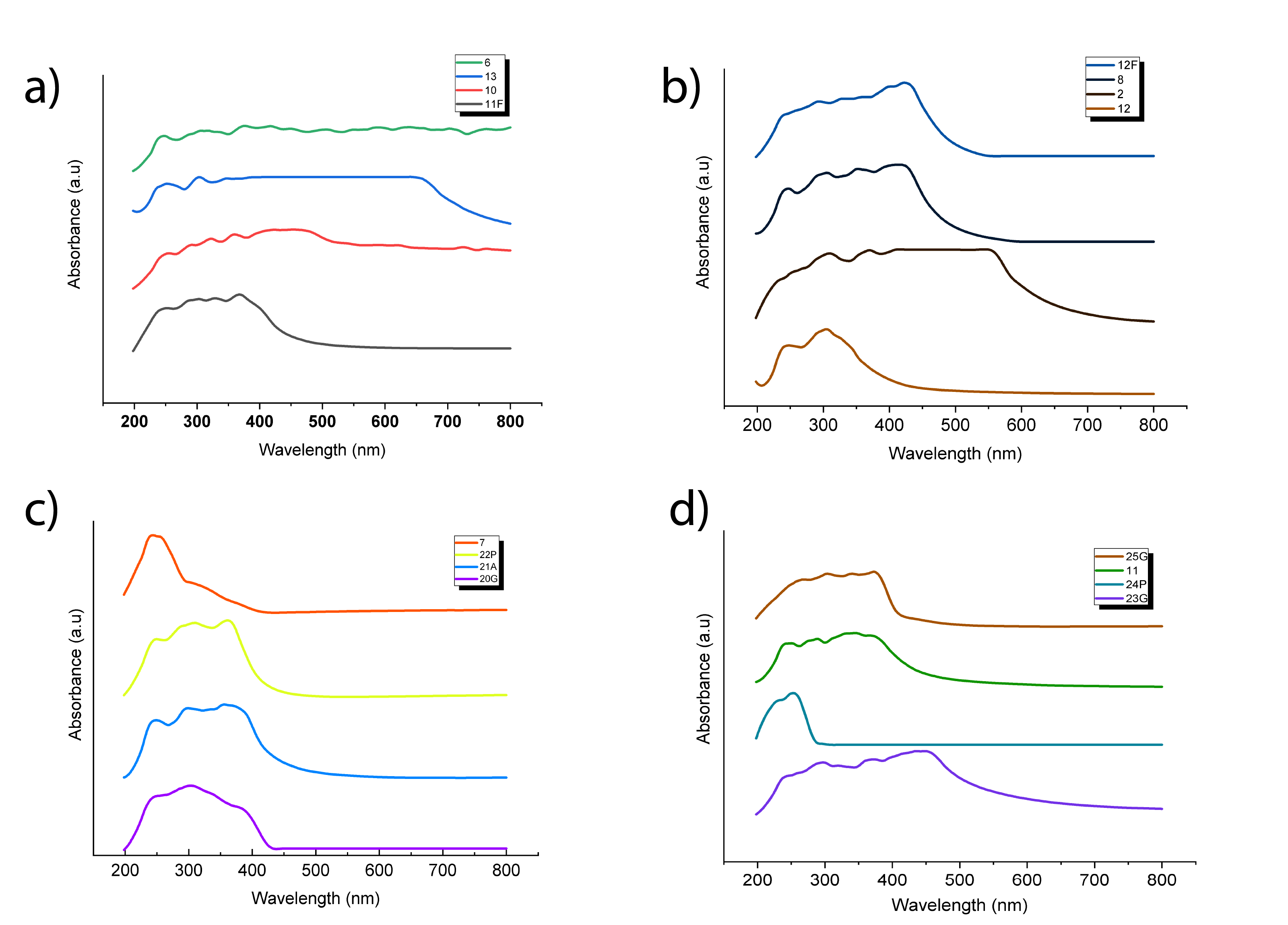}
\caption{The picture depicts the ultraviolet (UV) spectra of 16 distinct proteinoid molecules spanning from 198 to 800~nm in wavelength. Diverse absorbance readings across the spectrum suggest that several distinct proteinoid molecules are present in the sample.
}
\label{pppsssazx999}
\end{figure}

\begin{table}[!tbp]
\centering
 \caption{The maximum absorption wavelength $\lambda_{max}$ and the optical band gap  $E_{g}$ for each proteinoid. The experiment shows that the maximum absorption wavelength is the reverse of the optical band gap. This means that the proteinoids can absorb light at different wavelengths.}
 \label{table:1}
 \begin{tabular}{||c c c c ||} 
 \hline
   Proteinoid  & \textbf{$\lambda$ } (nm)  & $E_{g} $ (eV) &\textbf{Code of Exp.} \\ [0.5ex] 
 \hline\hline
 \textbf{L-Glu:L-Asp} & 553 & 2.33& 2  \\ 
 \textbf{L-Glu:L-Phe:L-His} & 423 & 2.93 & 8  \\
 \textbf{L-Lys:L-Phe:L-His} & 303 & 4.10  & 13 \\
 \textbf{L-Glu:L-Phe} & 421 & 2.95 &10   \\
 \textbf{L-Glu:L-Asp:L-Lys} & 343 & 3.62  & 11 \\
 \textbf{L-Glu} & 302 & 4.11 &12 \\
 \textbf{L-Glu: L-Phe:PLLA} & 367 &3.38  & 11F \\
 \textbf{L-Lys:L-Phe-L-His:PLLA} & 427 &2.90  & 12F \\
 \textbf{L-Lys:L-Phe:L-Gly} & 243 & 5.10 & 7\\
 \textbf{L-Glu:L-Phe}& 375 & 3.31 &25G\\
 \textbf{L-Asp}& 358 & 3.46 &21A\\
 \textbf{L-Phe:L-Lys}&366 & 3.39& 22P\\
 \textbf{L-Glu:L-Arg}& 300& 4.13 & 20G\\
 \textbf{L-Phe}& 254& 4.88 &24P\\
 \textbf{L-Glu:L-Asp:L-Pro}& 458 & 2.71 &23G\\
 \textbf{L-Glu:L-Asp:L-Phe} & 374  & 3.32 & 6 \\ [1ex] 
 \hline
 \end{tabular}
\end{table}

Sixteen distinct proteinoids --- determined by various combinations of amino-acids --- were analysed via their UV spectra 
(Fig.~\ref{pppsssazx999}). The spectrum was measured between 198 and 800~nm. Each proteinoid was discovered to have its own spectral fingerprint (Tab.~\ref{table:1}. The presence of aromatic amino acids in the proteinoids was shown by the presence of a peak in the spectra near 280~nm.

What is known as a material's optical band gap is the energy difference between its highest valence band and its lowest conduction band. As such, it is a key factor in determining how a material will behave electrically and optically. It can be used to determine whether a material is an insulator or a semiconductor, and it can also be used to foretell whether or not a certain material will absorb or emit light. The approximation of a material's optical band gap is calculated using an equation using the energy band gap ($E_{g}$), the Planck constant ($h$), the speed of light ($c$), and the wavelength ($\lambda$)~\cite{kheirabadi2022learning}.
\begin{equation} \label{eq202}
E_{g}=\frac{hc}{\lambda}=\frac{1240}{\lambda}
\end{equation}

Several absorption peaks can be seen in proteinoids, as seen in Fig.~\ref{pppsssazx999} This is because of the molecular structure, which makes it possible for distinct parts to absorb light of different colours. 



\subsection{Voltage-Induced Conductive Pathways}


\begin{figure}[!tbp]
\centering
\includegraphics[width=0.99\textwidth]{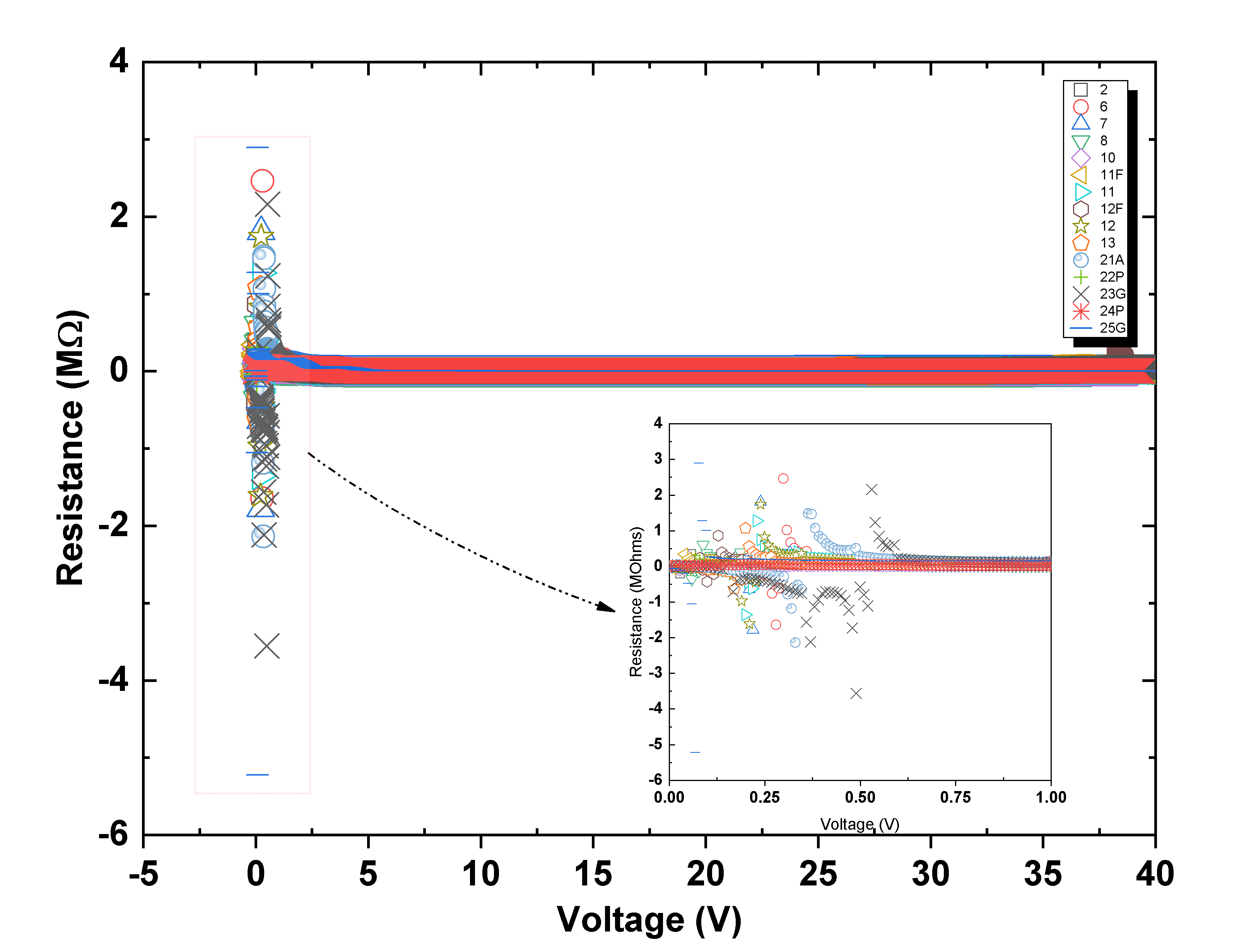}
\caption{Resistance versus voltage measurements of 16 types of proteinoids. The data shows a wide range of resistance values across different proteinoids in response to increasing voltage.}
\label{ppsoiunpsssazx999}
\end{figure}

We apply a DC voltage between 0 and 3~V and measure the resulting change in resistance, denoted by $R_{1}$. Once this is done, the sample's resistance, represented by $R_{2}$, can be calculated. Typically, a lower result for $R_{2}$ than $R_{1}$ indicates that the sample's resistance has decreased. It is necessary to repeat this method for a variety of proteinoids to get an accurate reading of their resistance levels.

Taking into account voltage 0.08~V and resistance 2.9 M$\Omega$, the proteinoid L-Glu:L-Phe (code 10) appears to have the highest resistance of any proteinoid. This conclusion is supported by the data presented in Fig.~\ref{ppsoiunpsssazx999}. The graph demonstrates that L-Glu:L-Phe is more resistant to a given voltage than the other proteinoids. If compared to the resistance of other proteinoids, which is between 0.0067 and 2.464 M$\Omega$, this is a huge increase.

Surprisingly high resistance is seen in proteinoids at near-zero voltages, but this resistance drops exponentially with increasing voltage up to around 40~V, where it then plateaus.

As can be observed in Fig.~\ref{ppsoiunpsssazx999}, at low voltages proteinoids occasionally show a negative resistance, meaning that an increase in voltage results in a drop in current. 

The resistance of proteinoids increases when voltage between 0 and 0.25 or 0.5~V applied but starts to increase when the applied voltage exceeds 5~V, as seen in supplementary material. The expansion rate of L-Asp is the highest (code 21A). The ``Proteinoids Threshold'' describes this phenomenon: Electrostatic interactions between charged peptide units in proteinoids polypeptides are responsible for the proteinoids threshold. As the peptides establish stronger inter-molecular connections at low voltages, the resistance of the proteinoids rises. The energy needed to force the current through the proteinoid rises in proportion to the square of the increase in resistance. Yet, as the voltage increases, the electrostatic connections lessen, resulting in a decrease in the proteinoid's resistance. As a result of the reduced resistance, less energy is needed to drive the current through the proteinoid~\cite{wuthrich2014micromachining}.

\subsection{
``Forgetting": Resistance Rises In Absence of the Stimuli}

Here, we investigate the potential application of ``forgetting'' in unconventional computing by observing how turning off stimulation alters the resistance along the proteinoid conductive channel. The outcomes of certain computations, e.g. machine learning, genetic algorithms, evolutionary computing, can be affected by the ability to ``forget".

Our methodology for analysing the findings involves taking readings of the resistance of the conductive pathways at regular intervals and comparing them to the readings taken before the stimulation began. This information is used to assess memory loss --- represented as a conductivity decline --- across a range of proteinoids. 

To visualise how a device's current ($I$) and voltage ($V$) are related, an $I-V$ curve can be drawn. The parameters that can be derived from an $I-V$ curve vary with the component or device being studied. For instance, the slope of an $I-V$ graph can be used to determine a resistor's conductance. Finding the point where the diode's $I-V$ curve begins to grow sharply allows one to calculate the diode's threshold voltage. 

The theory of varistors is fundamental to the investigation of proteinoids conductivity and other electrical phenomena. It asserts that the amount of current flowing through a material is equal to the ratio of the applied voltage ($V$) and a constant, $K_{0}$. The law (Eq.~\ref{eqpokm})

\begin{equation} \label{eqpokm}
I = \frac{K_{0}}{V^{\alpha}}
\end{equation}
expresses this connection between these two variables.
In order to comprehend how substances react when subjected to an electric field, the Varistor Theory is a crucial idea. 
The extent to which the electrical characteristics of a material change in response to a change in voltage is described by this relationship (Eq.~\ref{eqpokm}). The material's intrinsic constant $K_{0}$ characterises the level of its non-linearity.

Many different types of materials, including semiconductors, ferroelectrics, and insulators, are predicted by the Varistor Theory. Semiconductors are widely employed in electronics because of their malleability in electrical applications. The resistance of these components changes in response to an external voltage. Materials with a permanent electric dipole moment are known as ferroelectric materials and have several practical uses. Insulators prevent electrical current from flowing because their composition prevents electrons from freely moving~\cite{levine1975theory}. Using the Eq.~\ref{eqpokmklo}, we can calculate the stimulation length in seconds (See Supplementary Material section).
\begin{equation} \label{eqpokmklo}
\text{Stimulus length}= \frac{\text{Voltage Range}}{\text{Each Stimulus Duration}}
\end{equation}
with the voltage range being 3 V. The time of dropping resistance measures how long it takes for the electrical simulation to reach its peak resistance before it starts to drop.

\begin{table}[!tbp]
\centering
\caption{Quantiles of Resistance distribution in $\Omega$s ($\Omega$) for 16 different proteinoids.}
\begin{tabular}{|c|c|c|c|c|c|c|}
	\hline
	  & R & R & R & R &R & R  \\
        Code             & 100\%   & 99.5\%  & 75.0\% & 50\% & 25.0\% & 0.0\%  \\
    of Exp. & maximum &  & quartile  &median  &quartile  & minimum \\
     & ($\Omega$) & ($\Omega$) & ($\Omega$) & ($\Omega$)& ($\Omega$) &($\Omega$) \\
     & & & & & &\\
	\hline\hline
	2 & 62084437 &  4359.9654  & 474.32786  & 7.1971977  & $-$2785.324 &  $-$9984000 \\
	\hline
	 6 &  416544.06 & 6934.5387 & 1.2881781  & 0.1536745 & $-$26.63374  & $-$7503421 \\
	\hline
	7 &362430464  &  10110.566&261.35351  & 217.14726 & $-$242.3689 & -7.71E+12 \\
	\hline
	8 & 76091.049  & 6183.5833&2325.6716  &1418.6132   & 692.87616 & 84.70032 \\
	\hline
	10 & 4404.9648 & 2688.7351 & 1628.779 &  1129.6275 &554.34378 & 112.67969 \\
	\hline
	11 & 303036.25 & 8691.1443 & 1391.5009 & 946.35925  & 628.69757   & $-$431974.8 \\
	\hline
	11F & 7.411E+12  & 24619418 &  586.01653 &  195.45464 &$-$4082.635  &$-$1.56E+13 \\
	\hline
	12F & 7186565.8 & 107251.93 & 2041.8851   &  741.79247 &$-$2451.287&  $-$7331591  \\
	\hline
         12  &      25315922       &  286739.73      &   9459.7533         &5205.7754   &2926.6472  & $-$66917508\\
           \hline
         13   &    1801.1524         &  1630.706      &   831.3615         & 430.28272  & 202.04338 & 36.918475\\
           \hline
         20G   & 2.389E+11            &   149585.92     &   39.122431         & $-$0.157974  &$-$52.63574  & $-$2.14E+11\\
           \hline
          21A  &    148750.47         &   8159.6149     &  $-$392.6154          & $-$1801.791  &$-$3559.802  & $-$341609\\
           \hline
         22P   &     3239.7874        &  2331.8801     &   831.36735         &  391.21415 &261.61789  & 90.301161\\
           \hline
          23G  &       1061668.3      &  20931.568      &   3288.7011         & 1871.6031  & 376.94064 &$-$2409099 \\
           \hline
            24P  &     9.823E+10        & 18445.175       & 3.7424821           & 0.5117157  &$-$3.252959  &$-$2.07E+11 \\
           \hline
            25G  &  3638.8308           &2393.2651        & 1433.5912            &   740.98793&379.97653  &74.836721 \\
           \hline
	\end{tabular}
\label{hmllfnjbhhssshsnf}
\end{table}

The term ``quantile" is used to describe the equal division of a data set. The 99.5 percentile is the value below which 99.5\% of the data points lie; the 75.0 percentile is the value below which 75.0\% of the data points lie; and the 0.0 percentile is the lowest value in the dataset.

\begin{figure}[!tbp]
\centering
\includegraphics[width=1\textwidth]{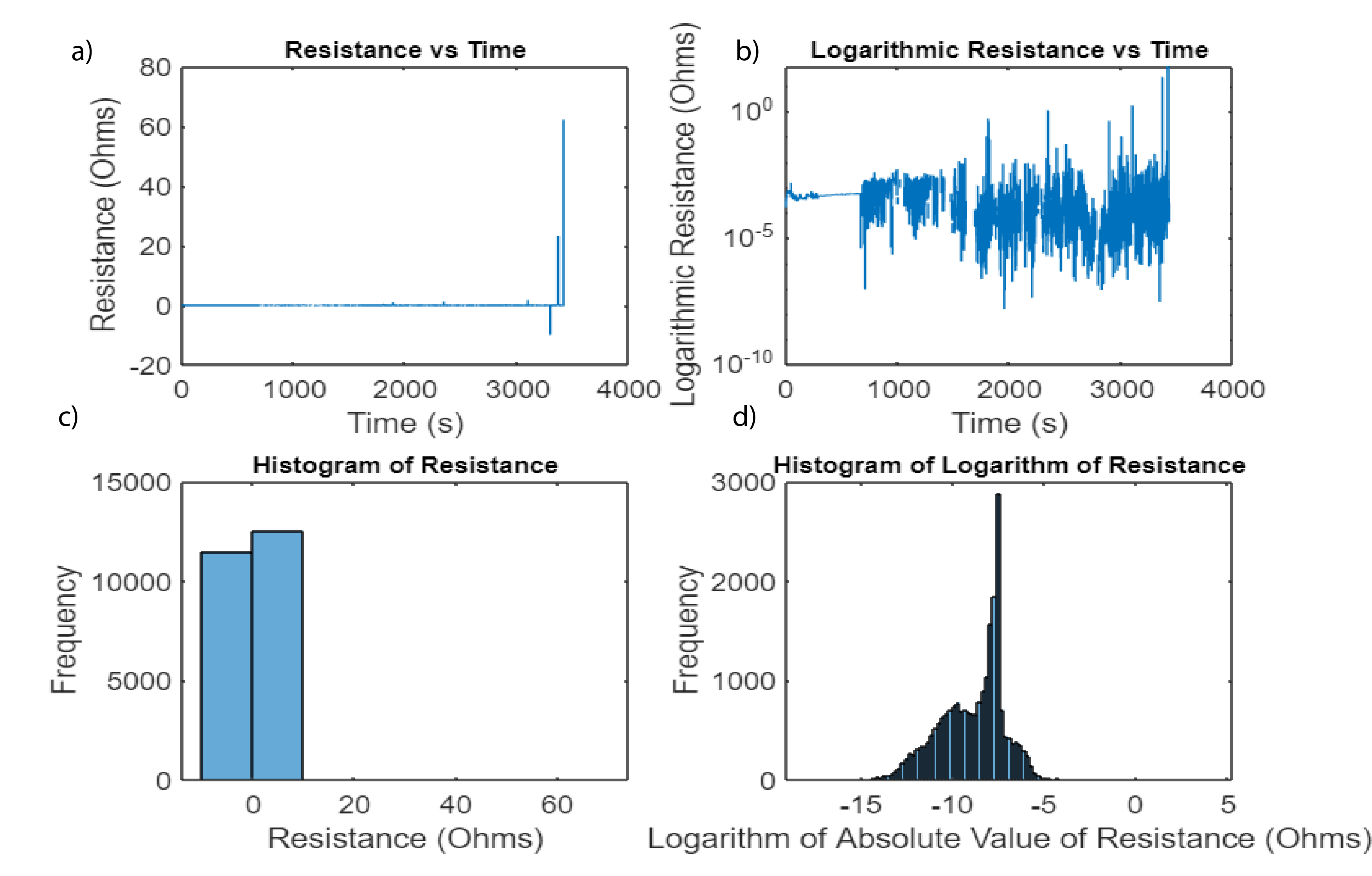}
\caption{This figure shows a),b) the resistance with code 2 over time and c),d) the distribution of resistance measurements. The upper 95\% mean of resistance is 9038.5 $\Omega$ and the lower 95\% mean is 1930.7 $\Omega$.}
\label{fggdfasfda;cm}
\end{figure}

Proteinoid 2, see codes of the proteinoid in Tab.~\ref{table:1}, resistance quantile measurements (Fig.~\ref{fggdfasfda;cm}) reveal some intriguing trends as follows. 100\% of the total resistance was reached at a value of 62.1 M$\Omega$. Although most resistance values were far less than the maximum value, the 99.5\% quantile value of 4360 $\Omega$ demonstrates this. When compared to the highest possible value, the median of 7.197~$\Omega$ is quite low. For a perfect percentage, the lowest value was $-$9984000 $\Omega$. These findings demonstrate that the resistance values of proteinoid 2 varied widely, with some extreme cases present. There appeared to be a fair deal of fluctuation in the strength of proteinoid 2, as the majority of resistance readings were significantly lower than the maximum.

Using a quantile regression analysis, the range of resistance values was calculated for proteinoid 6, with quantiles of 100\% 416544.1 $\Omega$, 99.50\% 6934.5 $\Omega$, 97.50\% 3984.6 $\Omega$, median 0.15 $\Omega$, and 0.01\% quantile $-$7503421 $\Omega$. According to the findings, the levels of resistance can be anywhere from $-$750.3 $\Omega$ to $+$41654.1 $\Omega$. 
The data also reveal that 416544.1 $\Omega$ is a significantly greater resistance value than the rest of the values in the range. This indicates that there is a very high resistance number that stands in stark contrast to the rest of the data.

 \begin{figure}[!tbp]
\centering
\includegraphics[width=1\textwidth]{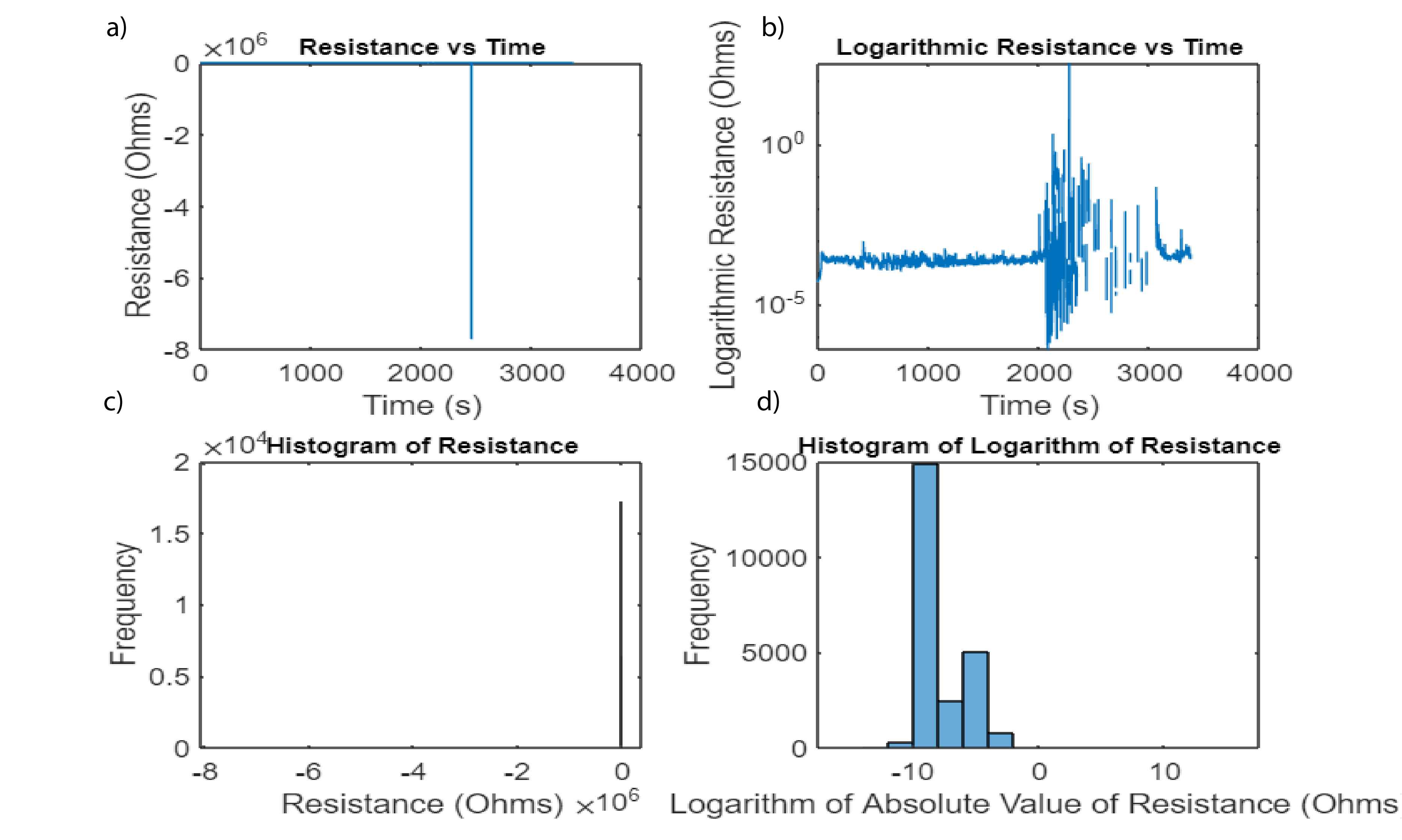}
\caption{ab) Resistance vs time from proteinoid 7 under constant potential 20 V and cd) histogram of resistance values with quantiles at 100\%,99.5\%,97.5\%,90\%,median, and 0.0\%. The resistance values in $\Omega$s are 362430464, 10111, 593.2, 339,261.4,217.1 and -7.7E+12, respectively. The graph of proteinoid 7 resistance versus time shows time passing in an arrow-like fashion. }
\label{Iwasthinki}
\end{figure}

Figure~\ref{Iwasthinki} shows that the 100\% percentile resistance value is 362 M$\Omega$. Resistance of 10111 $\Omega$ represents the 99.5\% quantile. The resistance value at the 97.5 percentile is 593.2 ohm, at the 90 percentile it is 339 ohm, and at the 75 percentile it is 261.4 ohm. There is a wide range of resistance values, with a median of 217.1 $\Omega$ and a lowest value of $-$7.7 $\times$ $\mathrm{10^{9}}$ M$\Omega$ (at the 0.0 percentile). This suggests a wide variation in resistance ratings for proteinoid 7 (see codes in Tab.~\ref{table:1}).

Compared to the other quantiles, the greatest resistance value of 362 M$\Omega$ is extremely high, suggesting that only a small fraction of the sample has such a high resistance value. The inclusion of a few samples with exceptionally low resistance values could possibly account for the fact that the lowest resistance value is much lower than the other quantiles, at $-$47.7E+9 M$\Omega$.

For proteinoids 11F,12F,13,22P,23G,25G and 11  (see codes in Tab.~\ref{table:1}),   an examination of the resistance as a function of time reveals several intriguing tendencies (See figures in Supplementary Material). The most striking pattern here is the sudden rise in resistance between zero and one hundred seconds. Following this boost, the resistance returns to a more manageable level and stays there until the 5000 second mark, at which point it begins to gradually grow again. Between 0 and 100 seconds, resistance increases dramatically, most likely because proteinoid microspheres are interacting with one another and forming more complex structures.This interaction makes the microspheres more robust to external influences, resulting in an increase in resistance. This boost in resistance is followed by a period of stability that lasts for around 5 minutes and a half. It's possible that the formation of increasingly complex structures has contributed to this steady rise in resistance.

Table~\ref{hmllfnjbhhssshsnf} demonstrates a large dispersion in resistance data histogram quantiles across 16 proteinoids. Proteinoid resistance was found to be maximal at the 100\% level, and to diminish with each succeeding quantile.
From the data of resistance of Table~\ref{hmllfnjbhhssshsnf}, we can see that the resistance range of protenoids 2,6,7,8,10,11,11F,12F,12,13,20G,21A,22P,23G, 24P and, 25G is between $-$9.98 M$\Omega$ and 62.1 M$\Omega$,$-$7.5 M$\Omega$ and 0.42 M$\Omega$, $-$7.71E+12 M$\Omega$ and 362.4 M$\Omega$,84.7 $\Omega$ and 0.76 M$\Omega$,112.7 $\Omega$ and 0.0044 M$\Omega$, $-$0.43 M$\Omega$ and 0.303 M$\Omega$, $-$7.33 M$\Omega$ and 7.2 M$\Omega$,$-$66.9 M$\Omega$ and 25.3 M$\Omega$, 36.9 $\Omega$ and 1801 $\Omega$, $-$2.14E+8 M$\Omega$ and 2.4E+8 M$\Omega$, $-$0.34 M$\Omega$ and 0.148 M$\Omega$, 90.3 $\Omega$ and 3239 $\Omega$, $-$2.4 M$\Omega$ and 1.06 M$\Omega$, $-$2.07E+8 M$\Omega$ and 9.8E+7 M$\Omega$, 74.8 $\Omega$ and 3638.8 $\Omega$ respectively. Increased resistance can be attributed to the proteinoid's more intricate structure, which features both polar and non-polar parts. Another cause for wide range of resistance values is due to the increased quantity of cross-linking between the microspheres, which increases the strength and stiffness of the proteinoid structure.

\begin{figure}[!tbp]
\centering
\includegraphics[width=1\textwidth]{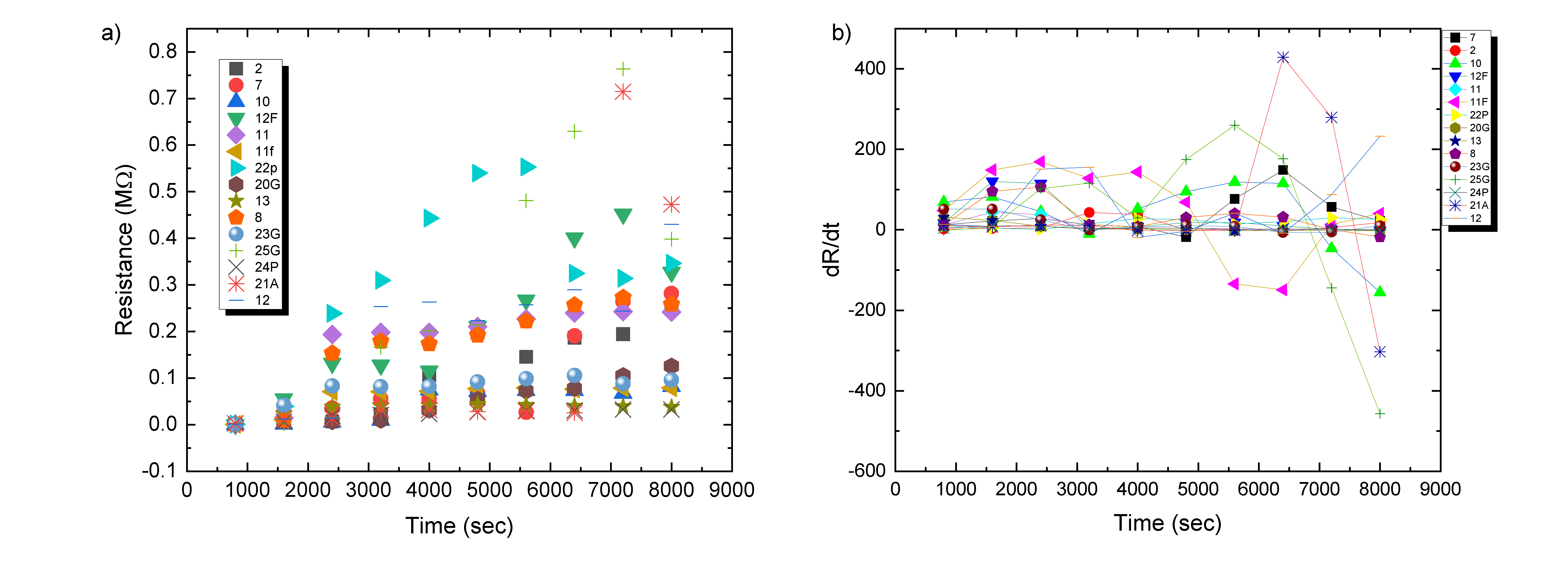}
\caption{The variation in resistance over time for 16 different proteinoids, codes of the proteinoids are given in Tab.~\ref{table:1}.
b) Resistance Derivation vs. Time.  This graph depicts the resistance rise over the time for 16 distinct proteinoids.
}
\label{gshcbahgjj}
\end{figure}

The results demonstrated that the 16 proteinoids' resistance increased with response to time (Fig.~\ref{gshcbahgjj}). Resistance of proteinoids varied from 0.00104 M$\Omega$ to 0.764 M$\Omega$ at the start, and peaked at 0.057 M$\Omega$ after 8000 seconds.

A first-order differential equation is one alternative mathematical model for describing the resistance's time-dependent evolution. One application of this type of equation is modelling the time-dependent resistance. Forms that the equation can take are:
\begin{equation} \label{eqphjjkkkokm}
dR/dt = k \times R + b
\end{equation}
Current versus voltage characteristics (see supplementary material) were measured 10 times at constant voltages between 0 $V$ and 40 $V$, and it was found that the slope of the I-V curve decreases with each additional measurement, indicating higher resistance as the number of measurements grows.

It is the material's physical qualities that are responsible for the drop in resistance. A rise in voltage energises the material's electrons, facilitating their freer movement and lowering the material's resistance. The I-V curve, which displays the resistance of a device, flattens out when more I-V readings are taken in a row, indicating an increase in resistance.

One possible explanation for the increased resistance is that the material itself was altered by the heating process. In response to an increase in voltage, a material's internal temperature rises, potentially resulting in a reduction in the material's resistance due to increased atomic vibration. To further illustrate the results, Fig.~\ref{gshcbahgjj} clearly demonstrates the resistance versus time increase for 16 different proteinoids.
Resistance at time $t$ is denoted by $R$, the rate of change of resistance over time is denoted by $k$, and the offset or bias term  is denoted by $b$. It is possible to estimate $k$ and $b$ using experimental data in order to fit the model to the observed behaviour.


\begin{figure}[!tbp]
\centering
\includegraphics[width=0.8\textwidth]{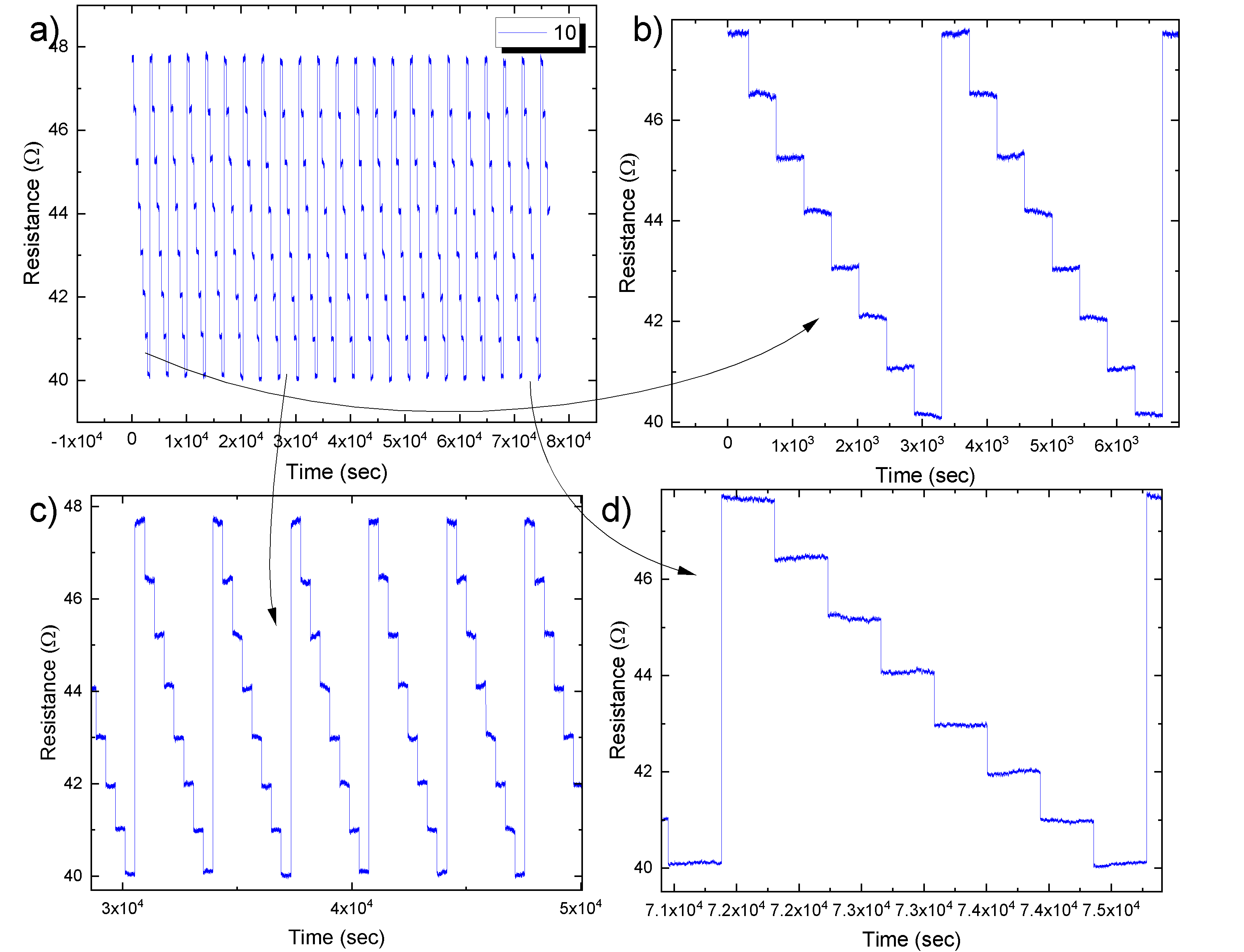}
\caption{This plot displays the resistance habituation with time for proteinoid 10, in response to a voltage arbitrary function with a frequency of 280~$\mu$Hz and an amplitude of 2~mV$_{pp}$.}
\label{kjvldsjhj;}
\end{figure}

Resistance adapted from 40.1 to 47.8 $\Omega$ when an arbitrary wave function with a frequency of 280 $uHz$ and an amplitude of 2 m$V_{pp}$ was applied, as depicted in Fig.~\ref{kjvldsjhj;}. Using a voltage function, the graph shows that the resistance of the material steadily increased from 40.1 to 47.8 $\Omega$, suggesting that it had become accustomed to the applied voltage.

\subsection{Habituation of Proteinoids Through ``Learning'' and ``Forgetting'' Cycles}

The habituation of proteinoids is studied by subjecting them to stimuli at timed intervals and then comparing pre- and post-stimulation resistance readings. This is accomplished by first measuring the proteinoids' resistance ($R$) without subjecting them to any stimulation. At times $T_{1}$, $T_{2}$, and so on, up to $T_{n}$, stimuli are introduced. The resistance of the proteinoids, designated $R_{i}$, is measured again after each stimulation. Habituation is demonstrated if and only if it can be shown that $R_{T_1}$ is smaller than $R_{T_2}$, and $R_{T_2}$ is less than $R_{T_3}$, and so on.

\begin{figure}[!tbp]
\centering
\includegraphics[width=0.8\textwidth]{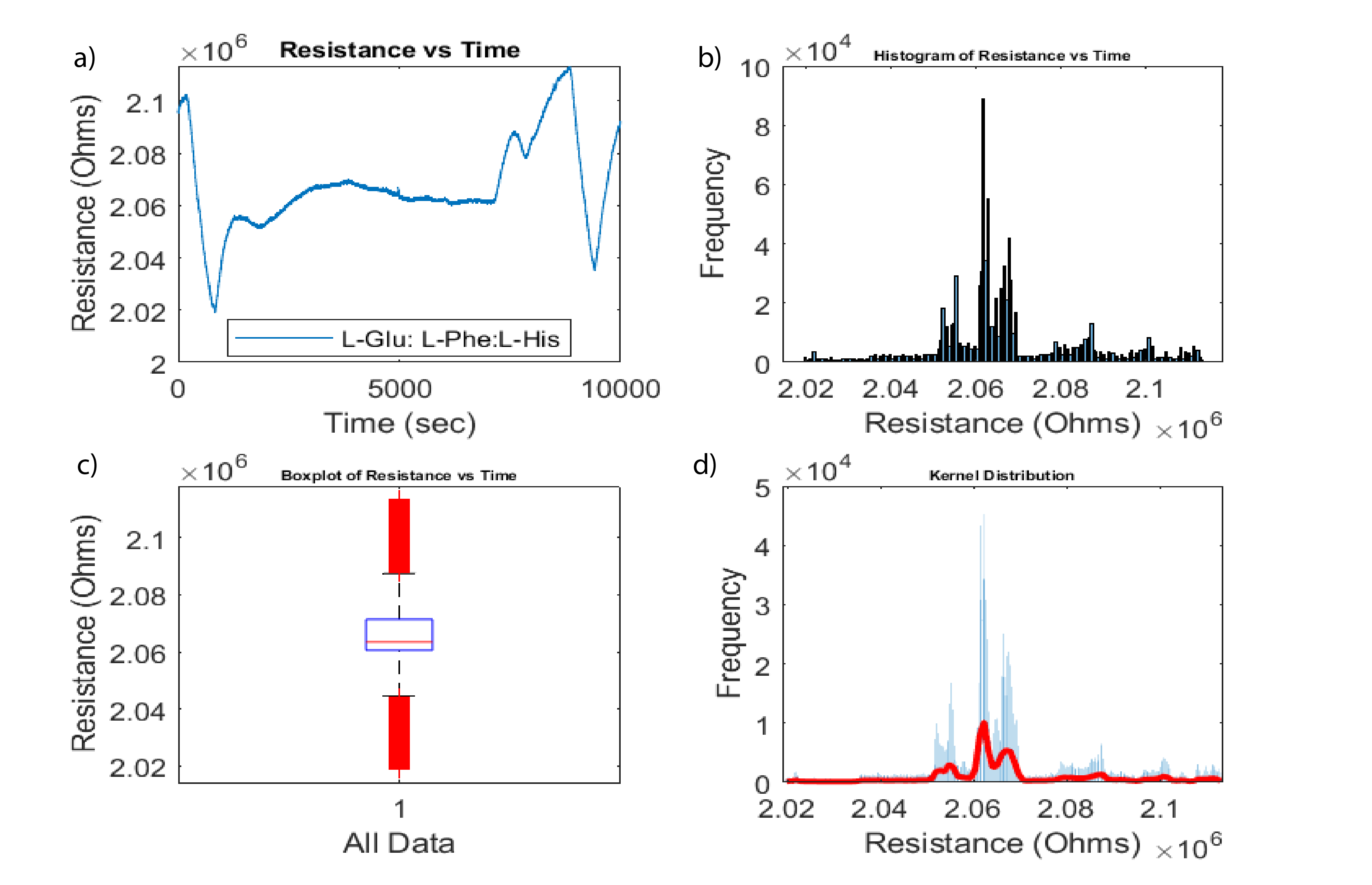}
\caption{a) Measurements of proteinoid code 8 resistance versus time using open circuit potentiometry.
 b) Distribution of Resistance levels for proteinoid 8 over time as shown by a histogram.
 c) Average resistance is found to be 2067342.3 $\Omega$, with a median of 2063572 $\Omega$ and a standard variation of 17364.2 $\Omega$, as shown by the Box plot. Inter-quartile range is 10812 $\Omega$.  d) Data-driven fitting of a Kernel distribution function.
}
\label{afgdsaghf}
\end{figure}

\begin{figure}[!tbp]
\centering
\includegraphics[width=0.8\textwidth]{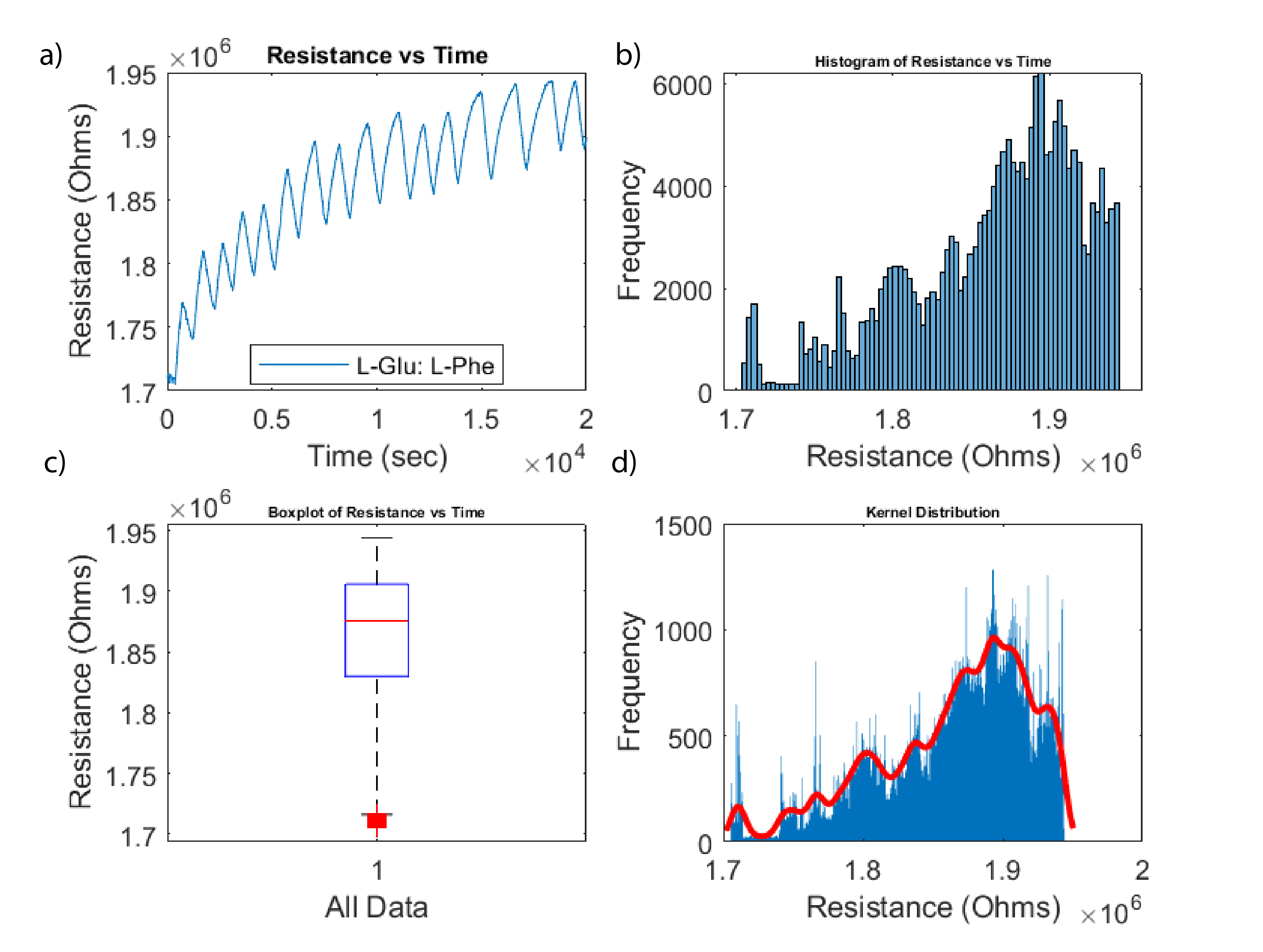}
\caption{a) The relationship between resistance and time was measured using open circuit potentiometry for proteinoid code 25G. b) ''Distribution of Resistance data for proteinoid 8 over time,'' a histogram showing the relationship between resistance and time. c) The Box plot identifies mean value of resistance 1863778.3 $\Omega$, median 1875762.2 $\Omega$, standard deviation 55364.1$\Omega$ and inter-quantile range 75956.5 $\Omega$ d) Data fit using the Kernel distribution function.
}
\label{gdsagashbdfhd}
\end{figure}

\begin{figure}[!tbp]
\centering
\includegraphics[width=0.8\textwidth]{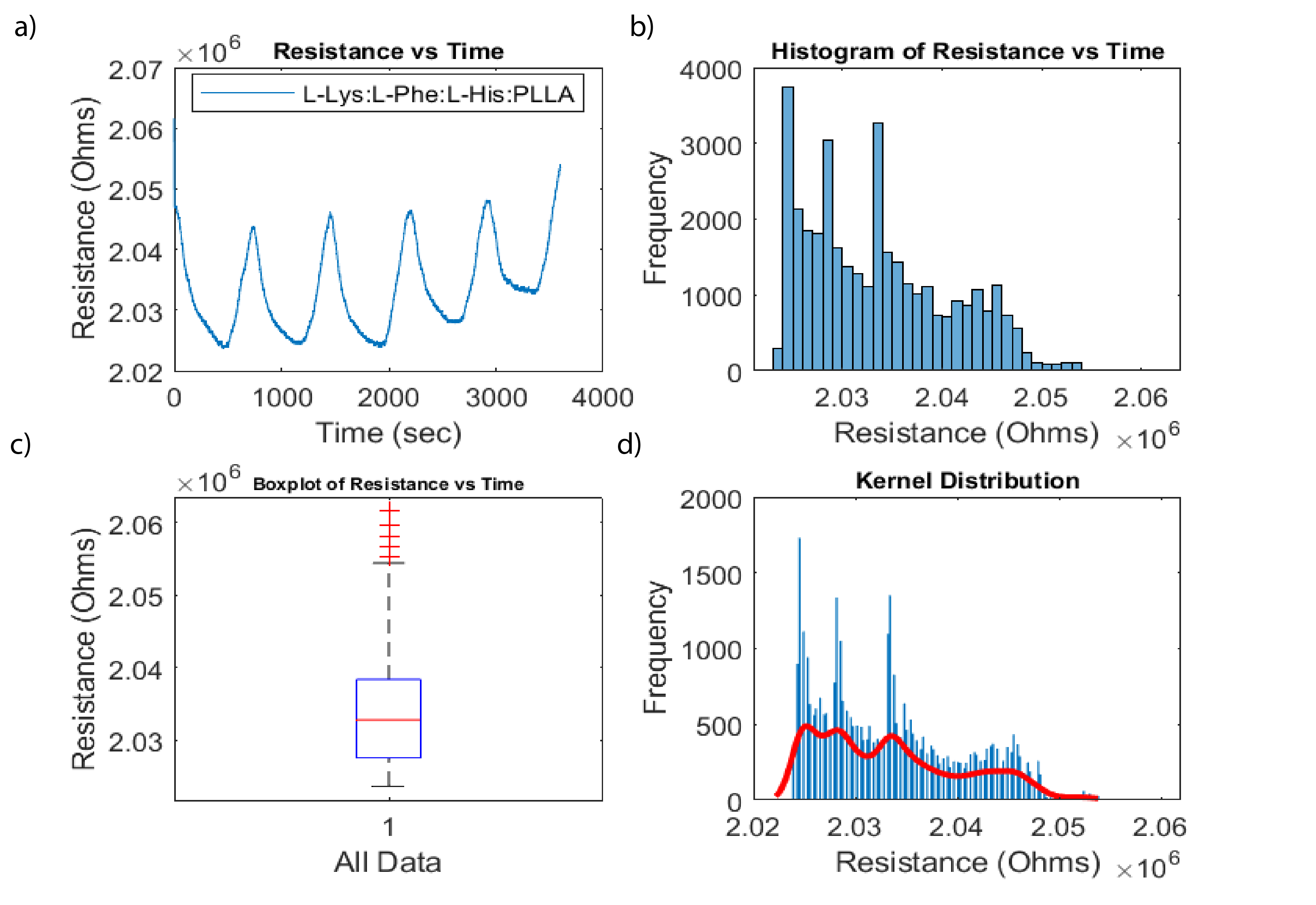}
\caption{a) The relationship between resistance and time was measured using open circuit potentiometry for proteinoid code 12F. b) ''Distribution of Resistance data for proteinoid 8 over time,'' a histogram showing the relationship between resistance and time. c) The Box plot identifies mean value of resistance 2033420.4982 $\Omega$, median 2032772 $\Omega$, standard deviation 7082.6487 $\Omega$ and inter-quantile range 10812 $\Omega$ d) Data fit using the Kernel distribution function.
}
\label{njincjedncdemlk}
\end{figure}

Proteinoids with codes 8, 25 G, and 12F (see codes in Tab.~\ref{table:1}) have been analysed for their habituation characteristics in Fig.~\ref{afgdsaghf}, \ref{gdsagashbdfhd}, and \ref{njincjedncdemlk}. According to the findings, proteinoid 25G has superior habituation characteristics compared to proteinoid 8. The average resistance of 25G proteinoids is 1863778.3 $\Omega$, the median is 1875762.2 $\Omega$, the standard deviation is 55364.1 $\Omega$, and the inter-quantile range is  75956.5 $\Omega$. Furthermore, there is a reasonable match between the results and the kernel distribution for proteinoid 25G. 

The resistance versus time data was fitted using the Kernel distribution, which is a nonparametric depiction of the probability density function (PDF). The likelihood of witnessing a specific resistance value at a specific time was faithfully represented by the Kernel distribution. This helped us make sense of our data and learn more about the correlation between resistance and time. 

The findings show that proteinoid 25G is better suited for uses like unconventional computing because it is less likely to cause habituation than proteinoid 8. The findings also imply that protenoid 25G could be used in other contexts, such as cell-based treatments, where resistance to habituation is desirable.

\section{Discussion}

In laboratory experiments we demonstrated that proteinoids can ``remember'', ``forget'', and ``habituate'' thus presenting essential features of living nervous systems.

Real-world memory is the recall of previously learned information. But, the ability to do this is not unique to humans; it can be recreated using biological molecules. Proteinoids can show signs of memory through a process called associative learning, which allows them to store and retrieve information. The proteinoids' ability to self-assemble and reproduce underlies their memory and retrieval capabilities. By experiencing a stimulus and responding to it in the future, a proteinoid can acquire the ability to learn. This property of proteinoids makes them suitable for use as memory in unconventional computing systems.

Forgetting is the process of erasing one's memories of the past through time. We have shown that proteinoids gradually lose the information --- in the form of resistance capacity~\cite{mougkogiannis2023transfer} --- when stimulation stops. This aids proteinoids' ability to focus on the most important data in unconventional computing by filtering out the noise. This makes it less likely that the proteinoids-based neuromorphic computing decices~\cite{adamatzky2021towards} will become overloaded with data.


\begin{figure}[!tbp]
\centering
\includegraphics[width=0.8\textwidth]{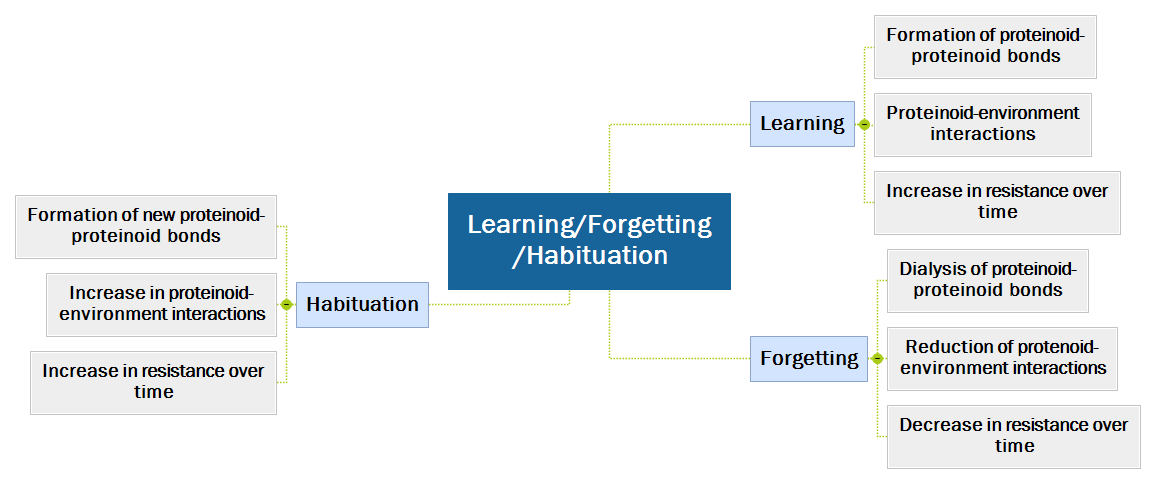}
\caption{This mind map demonstrates the mechanism of proteinoids' learning, forgetting, and habituation. Proteinoids learn by making new connections to external stimuli, forget by weakening and severing old connections, and get accustomed by learning to disregard extraneous stimuli. This process is exemplified by the fluctuation in resistance over time.
}
\label{vsdagbsaaasab}
\end{figure}

Fig.~\ref{vsdagbsaaasab} outlines the mechanism of learning, forgetting, and habituation, with a focus on the variation of proteinoid resistance across time.

Other non-trivial responses of proteinoids observed in experiments are discussed below.

In the experiments we found that for low voltages resistance becomes negative. This happens because of how proteinoids are built; under particular conditions, such low voltage, they can form complexes that cause this effect. Many proteinoid molecules connect with one another in these complexes, and as the voltage is raised, the proteinoids bind more securely to one another, thereby decreasing the current. Because of this drop in current, negative resistance is present. The use of bioelectronic devices is one area where negative resistance might be advantageous. With this quality, designers can craft oscillators and filters with very distinct electrical profiles, allowing them to execute very specialised functions. The decreased current draw of the negative resistance results in a lower power requirement for operation of the device, making this proteinoid feature useful for designing energy-efficient circuits.

We found that L-Glu:L-Phe is more resistant to a changing voltage than the other proteinoids.
This indicates that L-Glu:L-Phe is a better option for voltage regulation since it is more resistant to voltage at these levels. Several theories suggest that the structure, composition, and presence of particular functional groups in the L-Glu:L-Phe proteinoid are responsible for this enhanced resistance. For instance, in comparison to the other studied proteinoids, L-Glu:L-Phe has a larger proportion of hydrophobic moieties among its amino acid residues. The L-Glu:L-Phe proteinoid also has a larger concentration of carboxyl and amide groups, which may account for its enhanced resistance. Although it is folded into a $\alpha$-helical shape, this proteinoid is smaller than the others we tested. 

A high resistance was recorded in proteinoids at near-zero voltages, but this resistance drops exponentially with increasing voltage up to around 40~V. This discovery is crucial because it proves proteinoids may be used to build robust electrical circuits. As proteinoids can function at low voltages while still providing the required resistance, they make a great choice for designing dependable electric circuits. Additionally, proteinoids exhibit a consistent resistance even when the applied voltage exceeds 40~V. This opens the door for the usage of proteinoid circuits in many other contexts, including ones with varying voltage requirements.

As a result of their enhanced resistance, proteinoids can be used for a variety of purposes, including as electrical insulation, electrical shielding, and thermal insulation. The signal from biological systems can be detected and amplified using proteinoids because they are employed as a substrate in biosensors. The higher resistance of proteinoids is significant for these applications, since it helps to ensure that the signal is not degraded by electrical interference. In other words, the proteinoids' increased resistance to voltage is a direct result of their having gradually formed a more stable structure over time.

Organic, biocompatible, self-assembling, programmable, and responsive to external stimuli, proteinoids are ideal for use as unconventional computer materials. They are well suited for imitating neural networks and artificial brains due to their peculiar characteristics, such as action-potential-like spiking of electrical potential.

How can proteinoids use their electrical potential spikes to remember and retrieve information?

Primitive proteins, or proteinoids, may store information and retrieve it via electrical potential spikes. Information storage and processing in these organisms have been extensively explored because they provide such an intriguing paradigm for early life forms seen in prebiotic chemistry. Proteinoids have the power to construct memories through a variety of mechanisms, including memory formation and consolidation, adaption to changing environmental conditions, habituation and sensitivity to stimuli and forgetting rate and memory decay.

When it comes to proteinoids, how exactly do memories get stored in their microspheres brain?

Proteinoids use a wide variety of chemical processes and electrical potentials to build and consolidate memories. Using electrical potentials, proteinoids are able to store information within their structures, which in turn sets off a cascade of processes that results in the creation of memory. It is believed that electrostatic forces acting on the proteinoids are the mediating factor in the creation and consolidation of memories.

When faced with new stimuli and environments, how do proteinoids respond?

Proteinoids also possess the remarkable ability to adapt to novel stimuli and situations in their environment. Proteinoids can change their memory and learn from new ``experiences'' to adapt to their surroundings. Because of this, they are able to adapt to different stimuli and environments with relative ease and efficiency.

How fast do proteinoids ``forget'' things, and what causes their memories to fade over time?

A proteinoid's forgetting rate and memory loss can be affected by a number of variables, including as the stimulus's type, length, and electrical potential. While proteinoids can retain and retrieve information for a long time, the memory is more likely to fade the longer the stimulus last. The rate at which memories fade and are forgotten can also be affected by the voltage of the electrical potentials.

How can proteinoids show habituation and sensitisation to familiar and unfamiliar stimuli?

Habituation and sensitisation to novel or unfamiliar stimuli are also behaviours that proteinoids can display. It is possible to become more sensitive to a stimulus or to respond more strongly to it with repeated exposure, a process called sensitisation, as opposed to being accustomed to or desensitised to it, as is the case with habituation. Proteinoids rely on this sort of behaviour so that they can swiftly and accurately adapt to their surroundings.

Thus, employing proteinoids as memory devices in organic electronic unconventional computing devices has both positive and negative aspects. Proteinoids' ability to rapidly store and process massive amounts of data is a major benefit.

\section{Conclusions}

So far, the studies of proteinoids' cognitive abilities have shown encouraging results. Indeed, proteinoids have been shown to be capable of learning, memory, and forgetting. They have the ability to build conductive channels and microspheres, as well as the capacity to form memories and adaptation to environmental stimuli. This is a major development for the future of biomimetic and unconventional computing. More study and experimentation are needed, but it's possible that one day organic computer systems will mimic proteinoids' learning abilities. Ultimately, proteinoids have been shown to be an important and useful tool in the study of novel semiconductors, and they may be the key to enabling innovative approaches to unconventional computing.

\section*{Acknowledgement}

The research was supported by EPSRC Grant EP/W010887/1 ``Computing with proteinoids''. Authors are grateful to David Paton for helping with SEM imaging and to Neil Phillips for helping with instruments.


\end{document}